\documentclass[showpacs,preprintnumbers,amsmath,amssymb]{revtex4}

\usepackage{dcolumn}
\usepackage{bm}
\usepackage{indentfirst}
\usepackage{graphicx}

\def\Vec#1{{\bf #1}}

\newcommand{\p}{\perp}

\begin{document}

\preprint{USM-TH-198}

\title{Connection between the Sivers function and the anomalous magnetic moment}

\author{Zhun Lu}\email[Electronic address: ]{zhun.lu@usm.cl}
\author{Ivan
Schmidt}\email[Electronic address:
]{ivan.schmidt@usm.cl}\affiliation{Departamento de F\'\i sica,
Universidad T\'ecnica Federico Santa Mar\'\i a, Casilla 110-V,
Valpara\'\i so, Chile\\ and Center of Subatomic Physics, Valpara\'\i
so, Chile}

\begin{abstract}
The same light-front wave functions of the proton are involved in
both the anomalous magnetic moment of the nucleon and the Sivers
function. Using the diquark model, we derive a simple relation
between the anomalous magnetic moment and the Sivers function, which
should hold in general with good approximation. This relation can be
used to provide constraints on the Sivers single spin asymmetries
from the data on anomalous magnetic moments. Moreover, the relation
can be viewed as a direct connection between the quark orbital
angular momentum and the Sivers function.
\end{abstract}


\maketitle

The quark orbital angular momentum~\cite{seh74} (or quark transverse
motion) plays an important role for understanding the spin and quark
structure of the nucleon, since as shown by many
studies~\cite{emc88,jaf90,ji97a,hag98,ma98,har99}, it is the missing
block of the total nucleon spin. Also many interesting phenomena or
observables require the presence of quark orbital motion, among
which the Sivers function~\cite{sivers} has attracted a lot of
interest, since it is an essential piece in our understanding of the
single spin asymmetries (SSA) observed in semi-inclusive deeply
inelastic scattering (SIDIS). These SSAs have been measured recently
by both the HERMES~\cite{Airapetian:2004tw,hermes05} and
COMPASS~\cite{compass,compass06} Collaborations. Denoted as
$f_{1T}^{\perp}(x,\Vec k_\p^2)$, the Sivers function describes the
unpolarized distribution of the quark inside a transversely
polarized nucleon, which comes from a correlation of the nucleon
transverse spin and the quark transverse momentum. Although this is
a $T$-odd type correlation, it has been found that final state
interaction~\cite{bhs} (FSI) between the struck quark and the
spectator system can produce the necessary phase for a non-zero
Sivers function, and its QCD
definition~\cite{collins02,jy,belitsky,bmp03} has just been worked
out. Besides the single spin asymmetry, another important feature of
the Sivers function is that it encodes the parton's orbital angular
momentum ($L_z$) inside the nucleon. This comes from the fact that
the Sivers function requires the nucleon helicity to be flipped from
the initial to the final state, while the quark helicity remains
unchanged. A convenient tool to study this kind of single spin
asymmetry (or the corresponding Sivers function) is the light-front
formalism~\cite{bro98}, which can express the Sivers function as the
overlap integration of light-front wavefunctions differing by
$\Delta L_z=\pm 1$~\cite{bhs,bg06}. The same kind of overlap
integration~\cite{bl80,bhms,bdh} of light-front wavefunctions (with
$J_z=\pm 1/2$ in the initial and final states) also appears in the
anomalous magnetic moment of the nucleon, which apparently encodes
the quark orbital angular momentum~\cite{bhms}. Therefore, it is
interesting to find relations between the Sivers function and the
anomalous magnetic moment of the nucleon, which is the main goal of
this work. With such a relation one can constrain the Sivers
function and the related asymmetries from data on nucleon anomalous
magnetic moments, and viceversa. Also, the relation can be viewed as
a direct connection between the quark orbital angular momentum
distribution and the Sivers function.

The proton state can be expanded in a series of Fock states
$|n,x_iP^+,\mathbf{k}_\perp,\lambda_i\rangle$ with coefficients
$\psi_{n/p}(x_i,\mathbf{k}_\perp,\lambda_i)$, which are the
light-front wavefunctions of the proton:
\begin{equation}
\Psi_p(P^+,P^-,\mathbf{0}_\perp)=
\sum_n\psi_n(x_i,\mathbf{k}_\perp,\lambda_i)|n,x_iP^+,\mathbf{k}_\perp,\lambda_i\rangle.
\end{equation}
Here $x_i=\frac{k_i^+}{P^+}$ is the light-front momentum fraction of
the quark, $\lambda_i$ denotes the helicity, and
$\mathbf{k}_{\perp}$ its transverse momentum. The wavefunctions are
Lorentz-invariant functions of the kinematics of the constituents of
nucleon, $x_i$ and $\mathbf{k}_{\perp i}$, with $\sum_i^n x_i=1$ and
$\sum_i^n \mathbf{k}_{\perp i}=\mathbf{0}_\perp$.

As pointed out before, the Sivers function requires that the nucleon
wavefunctions in the initial and final state differ by one unit of
orbital angular momentum, and final state interactions play a
crucial role. It describes the interference of two amplitudes which
have different initial proton spin $J_z= \pm \frac{1}{2}$ but couple
to the same final state: Im$[M^*(J_z=+1/2 \rightarrow F)M(J_z=-1/2
\rightarrow F)]$. This can be realized by a gluon exchange between
the struck quark and the spectator system. There have been already
attempts~\cite{jmy02,bg06}, using the proton light-front
wavefunctions, to find a formula to calculate the Sivers functions.
In those papers the final state interaction phase needed for Sivers
functions has been incorporated into the wavefunctions. Another
possibility is to express the Sivers function as the product of
wavefunctions and the final state interactions term:
\begin{eqnarray}
f_{1T}^{\perp}& \propto & \ \sum_n\psi_{n}^{\uparrow
*}(x_i,\mathbf{k}_\perp,\lambda_i)G(x_i,x_i^\prime,\Vec
k_{\p,i},\Vec k_{\p,i}^\prime)
\psi_{n}^{\downarrow}(x_i^\prime,\mathbf{k}_\perp^\prime,\lambda_i^\prime),\label{f1twf}
\end{eqnarray}
where $G(x_i^\prime,\Vec k_{\p,i}^\prime,x_i,\Vec k_{\p,i})$ is the
final state interaction term, and the light-front wavefunctions in
this equation are the usual ones which do not contain the final
state interactions phase.

In the scalar diquark model, the calculation gets simplified and
some instructive result can be derived. In this model the kinematics
is determined by the momentum fraction and the transverse momentum
of the struck quark $(x,\Vec k_\p)$, since those of the spectator
diquark can be related to $(x,\Vec k_\p)$ by $x_D=1-x$,
$\mathbf{k}_{\perp D}=-\mathbf{k}_{\perp}$. The advantage of the
diquark model is that it is simple, and it works quite well in
describing other properties of the proton, such as the helicity
distribution~\cite{ma96}, the transversity~\cite{ma00} and the form
factors~\cite{ma02} of the proton. The light-front wavefunctions of
this two-body Fock state have the simple forms \cite{bl80,bhs}:
\begin{equation}
\left \{ \begin{array}{l} \psi^{\uparrow}_{+\frac{1}{2}} (x,{\Vec
k}_{\perp})=(M+\frac{m}{x})\, \varphi (x,{\Vec k}_{\perp})\ ,\\
\psi^{\uparrow}_{-\frac{1}{2}} (x,{\Vec k}_{\perp})=
-\frac{(+k^1+{\mathrm i} k^2)}{x }\, \varphi (x,{\Vec k}_{\perp})\ .
\end{array}
\right. \label{sdup}
\end{equation} for $J_z=+1/2$ and
\begin{equation}
\left \{ \begin{array}{l} \psi^{\downarrow}_{+\frac{1}{2}} (x,{\Vec
k}_{\perp})= \frac{(+k^1-{\mathrm i} k^2)}{x }\, \varphi (x,{\Vec
k}_{\perp})\ ,\\ \psi^{\downarrow}_{-\frac{1}{2}} (x,{\Vec
k}_{\perp})=(M+\frac{m}{x})\, \varphi (x,{\Vec k}_{\perp})\ .
\end{array}
\right. \label{sddown}
\end{equation}
for $J_z=-1/2$. Here
\begin{equation}
\varphi (x,{\Vec k}_{\perp}) = \frac{g/\sqrt{1-x}}{M^2-({\Vec
k}_{\perp}+m^2)/x-({\Vec k}_{\perp}+M_d^2)/(1-x)},
\end{equation}
and $M$, $m$ and $M_d$ are the masses of the proton, the quark and
the diquark, respectively.

\begin{figure}

\begin{center}
\includegraphics{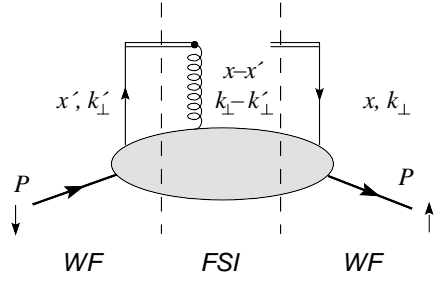}\caption{\small Diagram to calculate
the Sivers function. The arrows $\uparrow / \downarrow $ show the
polarizations of the nucleon.} \label{sivdg}
\end{center}

\end{figure}

According to Fig.~\ref{sivdg}, a formula to calculate the Sivers
function can be given by the overlap integration of the proton
light-front wavefunction:
\begin{eqnarray}
\hspace{-0.5cm}\frac{k^L}{2M}f_{1T}^{\perp
q}(x,\mathbf{k}_{\perp})&=&i\sum_n\int\frac{d^2k_\p^\prime
dx^\prime}{16\pi^3}\psi_n^{\uparrow\star}(x,\mathbf{k}_{\perp},\lambda_i
) G(x,x^\prime,\Vec k_\p,\Vec
k_\p^{\prime})\psi_n^{\downarrow}(x^\prime,\mathbf{k}_{\perp}^\prime,\lambda_i^\prime),\label{integral}
\end{eqnarray}
where $k^L=k^1-\textmd{i}k^2$, and $G$ is the kernel which contain
the final state interaction. In the scalar diquark model, the kernel
has a simple form~\cite{bhs,bh04}:
\begin{equation}
G(x,x^\prime,\Vec k_\p,\Vec
k_\p^{\prime})=\frac{iC_F\alpha_s\delta(x-x^\prime)}{2\pi((\Vec
k_\p-\Vec k_\p^\prime)^2+\lambda^2_g)},\label{kernel}
\end{equation}
which can be calculated in the eikonal approximation. Final state
interactions arise from gluonic exchanges between the struck quark
and the spectator system, which are necessary in order to insure
gauge invariance of the $\Vec k_\p$-dependent distributions. In our
case we use a one-gluon exchange (with momentum $k-k^\prime$)
approximation, which is commonly used in most model
calculations~\cite{bhs,jy,gg02,bbh03,yuan,bsy04,lm04a}. Here
$\lambda_g$ is the mass of the gluon, which is needed in general to
regularize the infrared divergence in the integration, while in our
case it can be set to $\lambda_g \to 0$. The sign in
Eq.~(\ref{kernel}) is according to the Trento conventions~\cite{tc}.
With Eqs.~(\ref{sdup}), (\ref{sddown}) and (\ref{integral}), one can
obtain the result for Sivers function that has given in
\cite{jy,bbh}. An expression similar to (\ref{integral}) has been
given in Ref.~\cite{bh04}, where the Sivers function (or the SSA) is
expressed as a product of a generalized parton distribution (GPD)
and a FSI term in impact parameter space, while our result is
expressed in momentum space, which can be connected with the result
in Ref.~\cite{bh04} by Fourier transformation.

\begin{figure}

\begin{center}
\includegraphics{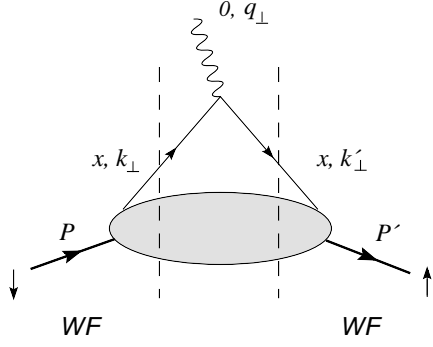}\caption{\small Diagram to calculate
the Pauli form factor (or anomalous magnetic moment). The arrows
$\uparrow / \downarrow $ show the polarizations of the nucleon.}
\label{ammdg}
\end{center}

\end{figure}

The same set of wavefunctions (spin-flipped in the initial and final
states) appearing in (\ref{f1twf}) also appears in the expression of
the Pauli form factor or anomalous magnetic momentum of the proton.
In light-front formalism the Pauli form factor is identified from
the helicity-flip vector current matrix elements of the $J^+$
current~\cite{bhms}
\begin{equation}
\left \langle P+q,\uparrow\left
|\frac{J^+(0)}{2P^+}\right|P,\downarrow \right
\rangle=-q^L\frac{F_2(q^2)}{2M}.
\end{equation}
Furthermore, from Fig.~\ref{ammdg}, one can express the Pauli form
factor in terms of the light-front wavefunctions as
\begin{eqnarray}
-q^L\frac{F_2(q^2)}{2M}&=&\sum_n\int\frac{d^2k_\perp
dx}{16\pi^3}\sum_j
e_j\psi_n^{\uparrow\star}(x_i,\mathbf{k}_\perp^\prime,\lambda_i)
\psi_n^\downarrow(x_i,\mathbf{k}_\perp,\lambda_i),\label{pff}
\end{eqnarray}
where
\begin{equation}
\mathbf{k}_{\perp}^\prime=\mathbf{k}_{\perp}+(1-x)\mathbf{q}_{\perp},
\end{equation}
here $q$ is momentum of the virtual photon, the momentum transfer
between the initial and final nucleon.

The anomalous magnetic moment is defined from the Pauli form factor
in the limit of $q=0$: $\kappa=(e/2M)F_2(0)$. After some algebra
calculation, the anomalous magnetic moment can be expressed in terms
of a local matrix element at zero momentum transfer:
\begin{eqnarray}
\hspace{-0.8cm}\frac{\kappa}{M}&=&-\sum_j e_j \sum_n\int
\frac{d^2k_\perp dx}{16\pi^3} \sum_{i\neq j}
\psi_n^{\uparrow\star}(x_i,\mathbf{k}_\perp,\lambda_i) x_i \left
(\frac{\partial}{\partial k_1}+\frac{i\partial}{\partial k_2} \right
) \psi_n^\downarrow (x_i,\mathbf{k}_\perp,\lambda_i).\label{amm}
\end{eqnarray}
From (\ref{f1twf}) and (\ref{amm}) we see that the same set of
light-front amplitudes, with the orbital angular momenta differing
by $\Delta L^z=\pm 1$ between the initial and final state, appears
in the calculation of the Sivers function and the anomalous magnetic
moment. Recently a study~\cite{burkardt06} has shown how the
non-vanishing anomalous magnetic moment constrains the quark orbital
angular momentum.

It is clear from the previous expressions that the relation we are
seeking between the Sivers function and the anomalous magnetic
moment will necessarily have to be approximate. This can be also
seen from the fact that the Sivers function, although leading twist,
has ($Q^2$) scale evolution dependence, which is not the case for
the anomalous magnetic moment.

In the scalar diquark model, the calculation of the anomalous
magnetic moment at the quark level has been given in \cite{bhms} as
\begin{eqnarray}
\frac{\kappa}{M}&=& \sum_{n=\pm 1/2} \int \frac{d^2k_\perp
dx}{16\pi^3} \psi_n^{\uparrow\star}(x,\mathbf{k}_\perp )(1-x)
 \left (\frac{\partial}{\partial
k_1}+\frac{i\partial}{\partial k_2} \right ) \psi_n^\downarrow
(x,\mathbf{k}_\perp).
\end{eqnarray}
Using the wavefunctions given in (\ref{sdup}) and (\ref{sddown}),
one gets the result
\begin{eqnarray}
\kappa&=&\frac{M g^2}{8 \pi^2} \int_0^1 dx
\frac{(1-x)^2(m+xM)}{(1-x)m^2+xm_d^2-x(1-x)M^2}\nonumber\\ &=&
\int_0^1 dx \frac{(1-x)A}{B},\label{ax}
\end{eqnarray}
where we have defined:
\begin{eqnarray}
A&=&\frac{g^2}{8\pi^2}(1-x)(xM+m);\\
B&=&(1-x)m^2+xM_d^2-x(1-x)M^2.
\end{eqnarray}

One can calculate the lowest $\Vec k_\p$-moment of $f_{1T}^{\perp,
q}(x,\Vec k_\p^2)$ defined as ($\lambda_g=0$)
\begin{equation}
f_{1T}^{\perp, q}(x) = \int d^2\Vec k_\p f_{1T}^{\perp, q}(x,\Vec
k_\p^2).
\end{equation}
According to Eq.~(\ref{integral}), which gives $f_{1T}^{\perp,
q}(x,\Vec k_\p^2)$,  and the wavefunctions given in (\ref{sdup}) and
(\ref{sddown}), we directly get
\begin{equation}
f_{1T}^{\perp, q}(x) = - \frac{\pi^2 C_F \alpha_s
A}{12B}.\label{f1tx}
\end{equation}

Defining
\begin{equation}
\kappa_q(x)=\frac{(1-x)A}{B}
\end{equation}
which satisfies the normalization condition
\begin{equation}
\int_0^1 dx \kappa_q(x)=\kappa_q,
\end{equation}
we arrive at a simply relation between $\kappa_q$ and
$f_{1T}^{\perp, q}(x)$ in the quark-diquark model:
\begin{equation}
 f_{1T}^{\perp, q}(x)=-\frac{\pi^2 C_F \alpha_s }{12
(1-x) }\kappa_q(x).\label{re1}
\end{equation}
The term $1/(1-x)$ in this equation is consistent with the result in
Ref.~\cite{bg06}. From this we can get:
\begin{eqnarray}
\int_0^1 dx (1-x) f_{1T}^{\perp, q}(x)&=& - \int_0^1 dx \frac{\pi^2
C_F \alpha_s }{12}\kappa_q(x) \nonumber \\ &=& - \frac{\pi^2 C_F
\alpha_s }{12}\kappa_q.\label{re2}
\end{eqnarray}
An approximate relation can also be obtained from the above
equation, as
\begin{eqnarray}
\langle 1-x \rangle F_{1T}^{\perp, q} &=& - \frac{\pi^2 C_F \alpha_s
}{12}\kappa_q,\label{re3}
\end{eqnarray}
where $F_{1T}^{\perp, q}$ is the first $x$-moment of $f_{1T}^{\perp,
q}(x)$ defined as $F_{1T}^{\perp, q} = \int_0^1 dx f_{1T}^{\perp,
q}(x)$, and $\langle 1-x \rangle$ is the average value of $(1-x)$ in
the proton. Therefore we obtain a interesting relation that provides
a constraint on the Sivers function from anomalous magnetic moment
data. It suggests that the moment of the Sivers function is
proportional to the anomalous magnetic moment contributed by the
same quark. Although this relation as been obtained in the diquark
model, it should hold in general with good approximation. This can
be seen from the general relations for the anomalous magnetic moment
and for the Sivers function in terms of light-front wavefunctions,
as given by Eqs.~(\ref{integral}) and (\ref{amm}), respectively.
Since both quantities calculated from the same wavefunctions, and
these are the ones that vary more strongly in the range of values
covered by the integrals, it is appropriate to apply the mean value
theorem, 
and therefore get the same type of relation as the one we have found
here. In this sense our approximate result can be considered more
general than the diquark model, and thus be applied to each quark
individually.

A more fundamental object than the anomalous magnetic moment is the
spin-flipped generalized parton distribution (GPD)
$E(x,\xi,t)$~\cite{dm,ji97a,ji97b,dvcs}. Actually the function
$\kappa_q(x)$ is $E(x,\xi, t)$ at the forward limit ($\xi=0$ and
$t=0$):
\begin{equation}
\kappa_q(x)= E_q(x,0,0).
\end{equation}
Thus in the scalar diquark model, $f_{1T}^{\perp, q}(x)$ is
proportional to $E_q(x,0,0)$. According to Ji's sum
rule~\cite{ji97a} ($J_q$ is the total angular momentum carried by
quark flavor $q$):
\begin{equation}
\int dx x \left (H_q(x, \xi ,t)+E_q(x, \xi ,t)\right
)=\frac{1}{2}J_q(t),
\end{equation}
which also holds in the forward limit. We see that the Sivers
function is related to the angular momentum of the parton inside the
nucleon, and therefore it is in fact sensitive to the orbital
angular momentum of the quark. There is then the possibility to get
information of the quark orbital angular momentum from the Sivers
functions.

Although the relation given in (\ref{re1}) and (\ref{re2}) is a
simple result based on the approximation of the scalar diquark
model, we can still apply the relation to given some prediction on
the Sivers asymmetry of the meson production in SIDIS processes,
such as the ratio of the asymmetries between different final mesons,
since in this case the model dependence is reduced. The Sivers
asymmetry is proportional to
\begin{equation}
A_{UT}^{Siv} \propto  \frac{ \langle \sum_a e_a^2 f_{1T}^{\p a}
D_1^a \rangle }{\langle \sum_a e_a^2 f_{1}^{ a} D_1^a \rangle },
\end{equation}
which can be extracted by the weighting function
$\sin(\phi-\phi_S)$, here $\phi$ and $\phi_S$ denote respectively
the azimuthal angles of the produced hadron and of the nucleon spin
polarization, with respect to the lepton scattering plane, $D_1^a$
is the unpolarized fragmentation function. We will focus on the
large $z$ regime that the valence quark contribution dominates, and
the disfavored fragmentation function can be ignored.

Since the Sivers function and the anomalous magnetic moment ``share"
the same set of the proton wavefunctions, as shown in
Figs.~\ref{sivdg} and \ref{ammdg}, one can start from the data of
the anomalous magnetic moment to provide constraints on the proton
wavefunctions, and then on the Sivers function. Similar methods have
been used in Ref.~\cite{bg06}, where a small Sivers asymmetry on a
deuteron target has been predicted, and Ref.~\cite{burkardt02},
where the sign of the Sivers asymmetries for different hadron
targets combining different fragmenting hadrons has been classified.

As figured out in Ref.~\cite{bg06}, the quark contribution dominates
over the gluon contribution in the case of Sivers
functions~\cite{Jacques}, which is also the result of an argument
based on large $N_c$ considerations~\cite{pobylista}. There are also
phenomenological supports of this conclusion from the SIDIS
experiment from COMPASS of CERN~\cite{compass}, as pointed out in
Ref.~\cite{bg06}, and the analysis on hadron production of $\pi$
given in Ref.~\cite{ans06}. Therefore in this work we only consider
the quark contribution to the Sivers functions and the corresponding
asymmetry.

One can constrain the proton wavefunctions by normalizing each $u$
and $d$ quarks contributions to the anomalous moments
$\kappa_p=1.79$, $\kappa_n=-1.91$. Isospin symmetry implies
\begin{eqnarray}
\kappa_{d/n}&=&\kappa_{u/p}, \\ \kappa_{u/n}&=&\kappa_{d/p}.
\end{eqnarray}
In the valence quark approximation we have:
\begin{eqnarray}
\kappa_p&=&(2)(2/3)\kappa_{u/p}+(-1/3)\kappa_{d/p}, \\
\kappa_{n}&=&(2)(-1/3)\kappa_{u/p}+(2/3)\kappa_{d/p}.
\end{eqnarray}
Thus one has $\kappa_{u/p}=0.835$, $\kappa_{d/p}=-2.03$. In the
following we use $\kappa_{u}$ and $\kappa_{d}$ to represent
$\kappa_{u/p}$ and $\kappa_{d/p}$, respectively.

Then we can write the ratio of the asymmetries between $\pi^+$ and
$\pi^-$ at large $z$:
\begin{equation}
\frac{A_{UT}^{Siv}(\pi^+)}{A_{UT}^{Siv}(\pi^-)} \approx \frac{2e_u^2
 f_{1T}^{\p u} D_1^{\pi^+/u} }{e_d^2
 f_{1T}^{\p d} D_1^{\pi^-/d}} \approx \frac{2e_u^2 \kappa_u}{e_d^2 \kappa_d} =-3.3.
\end{equation}
Also we have
\begin{eqnarray}
\frac{A_{UT}^{Siv}(\pi^0)}{A_{UT}^{Siv}(\pi^-)} &\approx&
\frac{2e_u^2
 f_{1T}^{\p u} D_1^{\pi^0/u}+e_d^2
 f_{1T}^{\p d} D_1^{\pi^0/d}  }{e_d^2
 f_{1T}^{\p d} D_1^{\pi^-/d}} \nonumber \\
 &\approx & \frac{2e_u^2\kappa_u+e_d^2 \kappa_d}{2e_d^2 \kappa_d} =
 -1.15,\\
\frac{A_{UT}^{Siv}(K^+)}{A_{UT}^{Siv}(K^0)} &\approx& \frac{2e_u^2
 f_{1T}^{\p u} D_1^{K^+/u} }{e_d^2
 f_{1T}^{\p d} D_1^{K^0/d}} \approx \frac{4e_u^2\kappa_u}{e_d^2 \kappa_d}
=-6.6.
\end{eqnarray}
For the above result we used isospin symmetry for the quark
fragmentation functions:
\begin{eqnarray}
D_1^{\pi^+/u}&=&D_1^{\pi^-/d}=2D_1^{\pi^0/u}=2D_1^{\pi^0/d},\\
D_1^{K^+/u}&=&2D_1^{K^0/d}.
\end{eqnarray}

The results show that in the large $z$ region, the Sivers asymmetry
of $\pi^+$ is 3 times larger than that of $\pi^-$ and with opposite
sign, which is consistent with the recent HERMES
result~\cite{hermes05} where at $z \sim 0.6$ a four times larger
asymmetry of $\pi^+$ is measured; the asymmetries of $\pi^0$ and
$\pi^-$ are similar in size; the asymmetry of $K^+$ is much larger
than that of $K^0$, nearly one order of magnitude; and
$A_{UT}^{Siv}(K^0) \sim 0$ since $\kappa_s=0$ in valence
approximation.

In summary, both the formula for calculating Sivers function and
that for calculating the anomalous magnetic moment of the proton,
can be expressed in terms of the same set of the light-front
wavefunctions, with helicity flipped from initial state to final
states. Using the overlap representations of both Sivers functions
as well as the anomalous magnetic moment, we give a simple relation
between these two observables, in the approximation of the scalar
diquark model. This relation is applied to provide constraints on
the Sivers single spin asymmetries in the valence regime from data
on anomalous magnetic moments. Also, the relation can be viewed as a
connection between the quark orbital angular momentum and the Sivers
function.

{\bf Acknowledgements.} This work is supported by Fondecyt (Chile)
under Project No.~3050047, and by the "Center of Subatomic Physics"
(Chile) .

\end{document}